\documentclass{emulateapj}
\slugcomment{to be submitted to ApJ}

\shorttitle{Disk Truncation and Overheated Region}
\begin{document}
\title{Formation of Overheated Regions and Truncated Disks around Black Holes;
Three-dimensional General Relativistic 
Radiation-magnetohydrodynamics Simulations}
\author{
Hiroyuki R. Takahashi\altaffilmark{1}, 
Ken Ohsuga\altaffilmark{2,3}
Tomohisa Kawashima\altaffilmark{1},
and
Yuichiro Sekiguchi\altaffilmark{4},
}

\altaffiltext{1}{Center for Computational Astrophysics, National
  Astronomical Observatory of Japan, Mitaka, Tokyo 181-8588, Japan}
\altaffiltext{2}{
Division of Theoretical Astronomy, National
  Astronomical Observatory of Japan, Mitaka, Tokyo 181-8588, Japan
}
\altaffiltext{3}{
School of Physical Sciences,Graduate University of
Advanced Study (SOKENDAI), Shonan Village, Hayama, Kanagawa 240-0193,
Japan
}
\altaffiltext{4}{
Department of Physics, Toho University, Funabashi, Chiba 274-8510, Japan
}

\begin{abstract}
Using three-dimensional general relativistic radiation magnetohydrodynamics simulations
of accretion flows around stellar mass black holes,
we report that the relatively cold disk 
($\gtrsim 10^{7}$K)
is truncated near the black hole.
Hot and less-dense regions, of which the gas temperature
is $ \gtrsim 10^9$K and more than ten times higher
than the radiation temperature (overheated regions), 
appear within the truncation radius.
The overheated regions also appear above as well as below the disk,
and sandwich the cold disk,
leading to the effective Compton upscattering.
The truncation radius is $\sim 30 r_{\rm g}$ 
for $\dot{M} \sim L_{\rm Edd}/c^2$,
where $r_{\rm g}, \dot M, L_\mathrm{Edd}, c$ 
are the gravitational radius, mass accretion rate,
Eddington luminosity, and light speed.
Our results are consistent with observations of 
very high state, whereby the truncated disk is thought to be embedded in
the hot rarefied regions.
The truncation radius shifts inward to 
$\sim 10 r_{\rm g}$ with increasing mass accretion rate
$\dot{M} \sim 100 L_{\rm Edd}/c^2$, which is very close to an innermost
stable circular orbit.
This model corresponds to the slim disk state observed in 
ultra luminous X-ray sources.
Although the overheated regions shrink if 
the Compton cooling effectively reduces the gas temperature,
the sandwich-structure does not disappear
at the range of $\dot{M} \lesssim 100L_{\rm Edd}/c^2$.
Our simulations also reveal that 
the gas temperature in the overheated regions
depends on black hole spin, which would be due to efficient energy 
tranpsport from black hole to disks through the Poynting flux, resulting
gas heating.
\end{abstract}

\section{Introduction}\label{intro}
It is widely believed that the black hole accretion flows 
are the central engine of the luminous compact objects like
active galactic nuclei and black hole binaries (BHBs).
The X-ray spectra of such objects are mainly composed of 
the soft component and the power-low component.
The soft component is dominant
over the power-low component in the high-soft state
as well as in the slim disk state,
and, in contrast, the power-low component is quite prominent
in the low-hard state and in the very high state 
\citep[see,][and references therein]{2007A&ARv..15....1D}.
The soft component is accepted to be multi-color disk blackbody,
which is emitted from the relatively cold, optically thick accretion disk.
On the other hand, 
the power-low component is thought to be produced due to 
the Compton upscattering in the hot and less-dense regions
(so-called disk corona).
From the observational point of view,
\cite{2006MNRAS.371.1216D} suggested the several models for 
the disk corona, and 
\cite{2004MNRAS.353..980K} reported the cold disk is truncated around the black hole
and the flow within the truncation radius is consist of the hot gas.
However, both the structure and 
the formation mechanism of the hot, rarefied plasma around the cold disk 
is not understood yet.

The most important mechanism to heat up the matter 
around the black hole is the dissipation of the magnetic energy.
Since a part of the kinetic energy of the accretion flow 
is converted to the magnetic energy via the magnetorotational instability,
the dissipation of the enhanced magnetic energy 
works to increase the gas temperature.
Indeed, this mechanism is thought to be origin of the viscous heating
in the accretion disk.
However, the radiative cooling prevents the gas from the rising in the
gas temperature in the dense regions.
Thus, study by the magnetohydrodynamics (MHD) is insufficient
and we need the radiation-MHD (RMHD).
The formation of the hot, rarefied regions above the cold disk 
has been reported by the RMHD simulations 
of a local patch of the disk \citep{2006ApJ...640..901H},
the global RMHD simulations \citep{2009PASJ...61L...7O,2011ApJ...736....2O}
and the general relativistic (GR) RMHD simulations 
\citep{2015MNRAS.454.2372S}.
However, the dependency of the mass accretion rate
hasn't been investigated well enough yet.

In addition, 
the rotation of the black hole might cooperate
the heating of the gas around the black hole.
The rotational energy of the black hole 
is extracted and transported outward by the Poynting flux.
If the transported electromagnetic energy is dissipated 
in the less dense regions,
the gas temperature would drastically increase.
This should be resolved by the GRRMHD simulations.

In this paper,
by performing the three-dimensional GRRMHD simulations,
we study the formation of the hot, rarefied regions around the 
black holes.
We will show that the relatively cold disk is truncated 
around the black hole and the flow becomes 
very hot within the truncation radius.
The hot, rarefied regions also appear above the cold disk.
In addition, we study the 
change of the size and the gas temperature 
of the hot regions 
due to the difference of the mass accretion rate and 
the black hole spin.
In \S 2, basic equations and numerical method are described. 
We present our results in \S 3.
In this section, we also discuss about the Compton cooling based on the
comparison between the cooling and dynamical timescales, since the
Compton cooling is not taken into account in our simulations. It is
reported by \cite{2015MNRAS.454.2372S} that the Compton cooling strongly
impacts on the gas temperature.
Finally \S 4 is devoted to summary and discussion.


\section{Basic Equations and Numerical Method}
In the present work, we numerically solve the GRRMHD equations. 
We hereafter take light speed $c$ as unity. 
The Greek suffixes indicate space-time components, and the Latin suffixes
indicates space components. 
The mass conservation equation is given by
\begin{equation}
 (\rho u^\nu)_{;\nu} = 0,\label{eq:masscons}
\end{equation}
where $\rho$ is the proper mass density, $u^\mu$ is the fluid four
velocity. 
The energy momentum conservation for the magnetofluid is given by
\begin{equation}
 (T^\nu_{\mu} + M^\nu_\mu)_{;\nu}= G_\mu, \label{eq:mhdenergymomentumcons}
\end{equation}
where $G_\mu$ is the radiation four force
(see equation [\ref{eq:rad_G}]).
The energy momentum tensor for fluid $T_\mu^\nu$ is given by
\begin{equation}
 T_\mu^\nu = \left(\rho + e + p_\mathrm{gas} \right)u_\mu u^\nu
  +p_\mathrm{gas} \delta_\mu^\nu,
  \label{eq:fluidenergymomentum}
\end{equation}
where $\delta_{\mu \nu}$ is the Kronecker delta,
$e$ is the gas internal energy, and 
$p_\mathrm{gas}$ is the gas pressure. 
Here we assume simple $\Gamma$-law for the equation of state and thus 
$e = (\Gamma-1)p_\mathrm{gas}$. 
We take $\Gamma=5/3$ in the following. 
The energy momentum tensor for electromagnetic field $M^{\mu \nu}$ is
given by
\begin{equation}
 M^{\mu \nu} = F^{\mu \alpha}F^{\nu}_{\alpha} 
  - \frac{1}{4}g^{\mu \nu}F^{\alpha \beta}F_{\alpha\beta},
  \label{eq:emenergymomentum}
\end{equation}
where $F^{\mu \nu}$ is the electromagnetic tensor and $g^{\mu \nu}$ is
the metric tensor. Here we absorbed factor $\sqrt{4\pi}$ into definition
of $F^{\mu \nu}$. 
In this paper, we adopted non resistive magnetohydrodynamics, so that
$F^{\mu \nu}$ satisfies
\begin{equation}
 u_\mu F^{\mu \nu} = 0.\label{eq:idealmhd}
\end{equation}

The magnetic four-vector is defined using magnetic field tensor $F^{\mu\nu}$
as
\begin{equation}
 b^\mu = 
\frac{1}{2}\epsilon^{\mu \nu \kappa \lambda}u_\nu F_{\kappa \lambda},
\label{eq:four_b}
\end{equation}
where $\epsilon^{\mu \nu \kappa \lambda}$ is the Levi-Chivita tensor. 
Substituting equation (\ref{eq:four_b}) into (\ref{eq:emenergymomentum})
gives more simple form;
\begin{equation}
 M^{\mu \nu} = b^2 u^\mu u^\nu + p_\mathrm{mag} g^{\mu \nu} - b^\mu b^\nu,
\end{equation}
where $p_\mathrm{mag}=b^2/2$ is the magnetic pressure. 
Since the components of $b^\mu$ are not independent, it is useful to
define magnetic field three vector $B^{i} \equiv F^{*it}$, where
$F^{*\mu \nu}$ is the dual of electromagnetic tensor. Then we obtain
\begin{eqnarray}
 b^t &=& B^i u^\mu g_{i\mu},\\
 b^i &=& \frac{B^i + b^t u^i}{u^t}.
\end{eqnarray}
Using these expressions, the induction equation $F^{*\mu \nu}_{;\nu}=0$
gives 
\begin{eqnarray}
 \partial_j (\sqrt{-g}B^j)= 0,\\
 \partial_t (\sqrt{-g}B^i)+\partial_j[\sqrt{-g}(b^i u^j - b^j u^i)]=0,
\end{eqnarray}
where $g=\mathrm{det}(g_{\mu\nu})$.

The energy momentum conservation for the radiation field is given by
\begin{equation}
 R^{\nu}_{\mu;\nu} = - G_\mu, \label{eq:radenergymomentumcons}
\end{equation}
where $R^\nu_\mu$ is the radiation energy momentum tensor.
In this paper, we employed M-1 formalism to close equations. 
Then, the radiation energy momentum tensor is given by
\begin{equation}
 R^{\mu \nu}= 4 p_\mathrm{rad} u^\mu_\mathrm{rad} u^\nu_\mathrm{rad}
   + p_\mathrm{rad} g_{\mu \nu},
\end{equation}
where $p_\mathrm{rad}$ is the radiation pressure and
$u^{\mu}_\mathrm{rad}$ is the radiation frame's four velocity
\citep{1984JQSRT..31..149L, 2013MNRAS.429.3533S, 2013PASJ...65...72K}.
The radiation four force $G^\mu$ is given by
\begin{equation}
 G^\mu= -\rho\kappa_\mathrm{abs}(R^\mu_\alpha u^\alpha + 4 \pi
  \mathrm{B}u^\mu)
  - \rho\kappa_\mathrm{sca}
  (R^\mu_\alpha u^\alpha + R^\alpha_\beta u_\alpha u^\beta u^\mu),
  \label{eq:rad_G}
\end{equation}
where $\kappa_\mathrm{abs}$ and $\kappa_\mathrm{sca}$ are the opacities for
absorption and scattering. We employ free-free emission/absorption 
and isotropic electron scattering,
\begin{eqnarray}
 \kappa_\mathrm{abs} 
  &=& 6.4\times 10^{22}\rho T_\mathrm{gas}^{-\frac{7}{2}}\ 
  \mathrm{cm^{2}\ g^{-1}}, \\
 \kappa_\mathrm{sca} &=& 0.4\ \mathrm{cm^{2}\ g^{-1}},
\end{eqnarray}
where $T_\mathrm{gas}$ is the gas temperature, which is 
related to the gas pressure as 
\begin{equation}
 p_\mathrm{gas} = \frac{\rho k_\mathrm{B} T_\mathrm{gas}}{\mu m_p}.
\end{equation}
Here $k_\mathrm{B}$ and $m_p$ are the Boltzmann constant and the proton
mass, and $\mu = 0.5$ is the mean molecular weight. 
The blackbody intensity is given by
$\mathrm{B}=a_\mathrm{rad} T_\mathrm{gas}^4/4\pi$, where $a_\mathrm{rad}$ is the radiation constant.

We solve these equations in Boyer-Lindquist polar coordinate
$(t,r,\theta,\phi)$ in Kerr-Schild space-time with black hole mass
$M_\mathrm{BH}=10 M_\odot$, where $M_\odot$ is the solar mass. Numerical
grid points are ($N_r, N_\theta, N_\phi$)=($264, 264, 64$) and
computational domain consists of $r=[r_\mathrm{H}, 250r_\mathrm{g}],
\theta=[0,\pi]$, and $\phi=[0,2\pi]$, 
where $r_\mathrm{H}$ is the horizon radius
and $r_\mathrm{g}=G M_\mathrm{BH}$ is the
gravitational radius. 
The radial grid size exponentially increases with radius. The 
$\theta$ is given by $\theta=\pi x_2 + (1-h)\sin(2\pi x_2)/2$ where
$h=0.2$ and $x_2$ is the uniform grid between $0$ and 1
\citep{2003ApJ...589..444G}. 
We adopted outflow boundary conditions at inner and outer radial
boundaries ($r=r_{\rm H}$ and $250r_{\rm g}$).
Reflective boundary condition is adopted at the
polar axis ($\theta=0$ and $\pi$).

The advection term is solved explicitly using Lax-Friedrich method,
while the source term describing interaction
between the gas and the radiation is integrated implicitly
\citep{2012MNRAS.426.1613R, 
2013ApJ...764..122T, 2013ApJ...772..127T,
2014MNRAS.441.3177M,
2014MNRAS.439..503S}. 
The divergence-free condition for the magnetic
field is satisfied by applying Flux-CT method
\citep{2000JCoPh.161..605T}. 

We start simulation from the equilibrium torus given by
\cite{1976ApJ...207..962F}. 
The inner edge of the torus is situated at
$r=20r_\mathrm{g}$, while the radius where the pressure has its maximum
value is $33r_\mathrm{g}$.
The initial torus is not in local thermodynamic equilibrium 
($T_\mathrm{gas}\neq T_\mathrm{gas}$), but we set small radiation energy
density $E_\mathrm{rad}=10^{-10}$ uniformly \citep[but,
see][]{2014MNRAS.439..503S}.  
Here $T_\mathrm{rad}=(\hat E_\mathrm{rad}/a_\mathrm{rad})^{1/4}$ is the
radiation temperature, and the hat denotes the quantity in comoving
frame.

The inner edge of the torus locates at
$r=20r_\mathrm{g}$ and there is a pressure maximum at 
$r=33 r_\mathrm{g}$. The maximum density of the torus $\rho_0$ is taken to be
a parameter. All physical quantities are normalized using $\rho_0$ 
and $r_\mathrm{g}$.
The time is normalized by $r_\mathrm{g}/c$, so that the unit time
is $\simeq 5\times 10^{-5}\ \mathrm{s}$ for $10 M_\odot$ black hole.
We embedded the weak poloidal magnetic field inside the torus. The
magnetic flux vector $A_\phi$ is given by 
$A_\phi \propto \rho$ and the ratio of the maximum $b^2$ and
$p_\mathrm{gas}$ at the initial
state is taken to be 100. 
In addition to the torus, we set the thin, unmagnetized hot
atmosphere. The density and gas pressure profile of the atmosphere are
given by $\rho=10^{-4} \rho_0 (r/r_\mathrm{g})^{-1.5}$ and
$p_\mathrm{gas}=10^{-6} \rho_0 (r/r_\mathrm{g})^{-2.5}$.

In this paper,
we perform three simulations.
We set $(\rho_0, a^*)=(10^{-4}\mathrm{g\ cm^{-3}}, 0)$ for run A, 
$(10^{-2}\mathrm{g\ cm^{-3}} , 0)$ for run B, 
and $(10^{-4}\mathrm{g\ cm^{-3}}, 0.9375)$ for run C.

\section{Results}
\subsection{overview of simulations}
\label{over}
In all simulations, 
the poloidal magnetic field lines in the torus 
begin to be twisted due to the differential rotation 
after the simulations start.
The toroidal component of the magnetic fields is enhanced.
The angular momentum is transported as 
the magnetorotational instability grows up inside the torus,
leading the mass accretion onto the black hole.
Then, the mass accretion rate suddenly increases
and the quasi-steady accretion disks are produced around the black hole. 

\begin{figure}[t]
 \includegraphics[width=8.5cm]{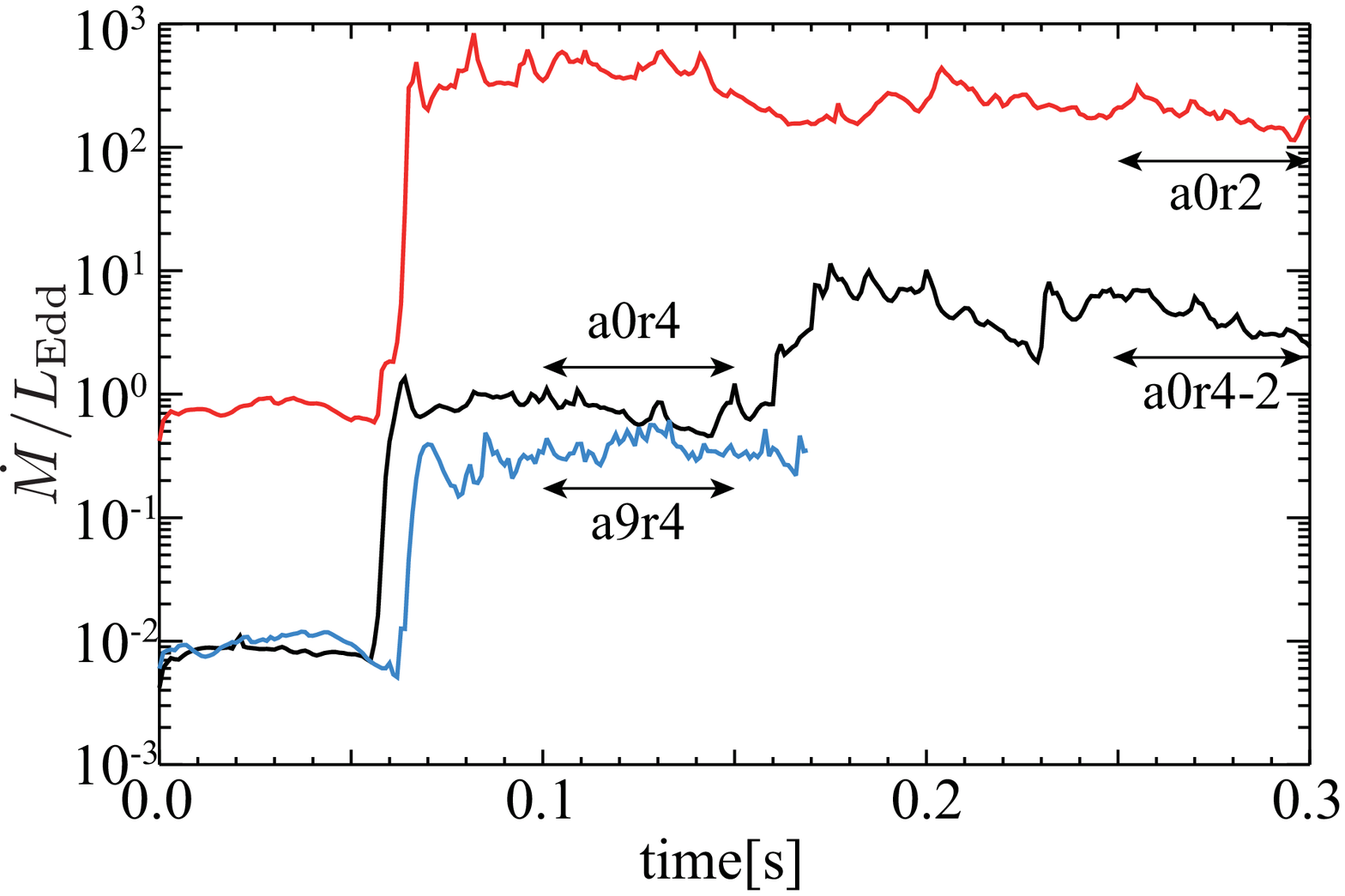}
  \caption{Time evolution of mass accretion rate. Black, red, and blue
 lines show results of runs A, B, and C, respectively.
 Arrows show the time interval for time average.}
\label{fig1}
\end{figure}

Figure \ref{fig1} shows the time evolution of the mass accretion rate 
$\dot{M}$. 
We find the rapid increase 
of the accretion rate at around $t=0.06\mathrm{s}$.
Then, a part of the initial torus reaches to the black hole.
At $t\gtrsim 0.06\mathrm{s}$, the accretion rate does not largely 
change,
although a transient amplification appears at $t\sim 0.17{\rm s}$
for run B (see black line).
Such an amplification is caused by the accretion
of the dense matter of the initial torus.

\begin{figure*}[t]
 \includegraphics[width=18cm]{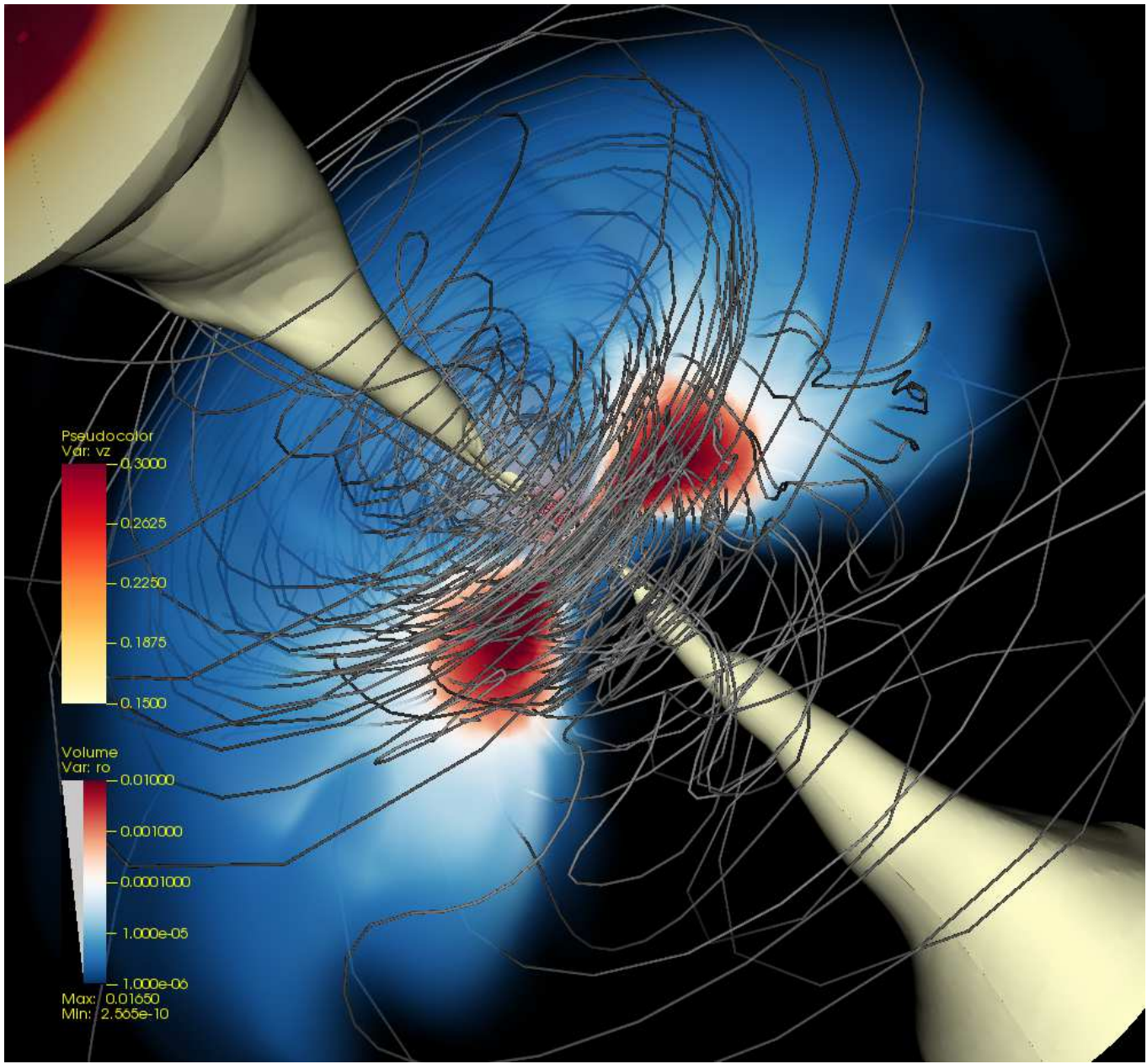}
  \caption{Global structure of radiation dominated accretion disks near
 the black hole at $t=0.3\ \mathrm{s}$ (run B). 
 The figure shows the density (blue-white-red volume rendering), the
 outflow velocity (white-red volume data), and the magnetic field
 lines (gray lines). }
\label{fig2}
\end{figure*}

As shown in this figure, 
the mass accretion rate for run A (red line)
is much larger than that for runs B and C,
since the larger $\rho_0$ is employed.
The accretion rate highly exceeds the critical rate,
$L_{\rm Edd}$, so that the radiation pressure-dominated disk 
is produced and the strong jets are launched from the disk surface
via the radiation force.
The overall structure of the accretion disk and jets at $t=0.3 \rm s$
is shown in Figure \ref{fig2}, where the disk is presented 
as blue-white-red volume rendering and the elongated white-red regions 
indicate the jets. 
Thin lines are magnetic field lines. We find 
that the toroidal magnetic fields are amplified inside the disk.
The disk-jet structure in this model is roughly consistent 
with that by \cite{2009PASJ...61L...7O, 2010PASJ...62L..43T, 2011ApJ...736....2O,
2014MNRAS.441.3177M,2015PASJ...67...60T,2015MNRAS.447...49S}.
For run B and run C,
the geometrical thickness of the disks is 
relatively small (see below), 
since the mass accretion rate is comparable to or
slightly smaller than the critical rate.

\subsection{structure of overheated region}
\label{structure}
We here take the time average 
between $t=[0.1, 0.15]\mathrm{s}$ for run C (model a9r4),
$t=[0.25, 0.3]\mathrm{s}$ for run A (model a0r2),
and  
$t=[0.1, 0.15]\mathrm{s}$ (model a0r4) 
as well as $t=[0.25, 0.3]\mathrm{s}$ (model a0r4-2)
for run B.
The time averaged mass accretion rate of model a0r4
($0.73L_\mathrm{E}$) is close to that of model a9r4
($0.38L_\mathrm{E}$),
so that we can discuss about the effect of the rotation of the black hole
by comparing these two models.
Here, we note that the inflow-outflow equilibrium is achieved 
within $r\sim 15 r_{\rm g}$ in two models (see, table 1).
On the other hand, 
we use models a0r2 and a0r4-2
in order to investigate the 
difference of the disk structure 
due to the difference of the mass accretion rate,
since the mass accretion rates are quite different,
$4.3L_\mathrm{E}$ (a0r4-2) and $430L_\mathrm{E}$ (a0r2).
In both models, the flow is in inflow-outflow equilibrium 
within $r\sim 20 r_{\rm g}$.

Here, we define following averages of
a physical quantity $f=f(t,r,\theta,\phi)$. 
The azimuthal average:
\begin{equation}
 <f>_{w,\phi} = 
  \frac{\int d\phi \sqrt{-g}f w}
  {\int_0^{2\pi} d\phi \sqrt{-g}w},
\end{equation}
and the azimuthal and polar average:
\begin{equation}
 <f>_{w,\theta \phi} = 
  \frac{\int d\phi d\theta \sqrt{-g}f w}
  {\int d\theta d\phi \sqrt{-g}w},
\end{equation}
where $w$ is a weight function. In this paper, we used a mass density as
a weight function ($<...>_{\rho}$), or we take an average without a
weight function ($<...>_{1}$). 
\begin{figure}[t]
 \includegraphics[width=8.5cm]{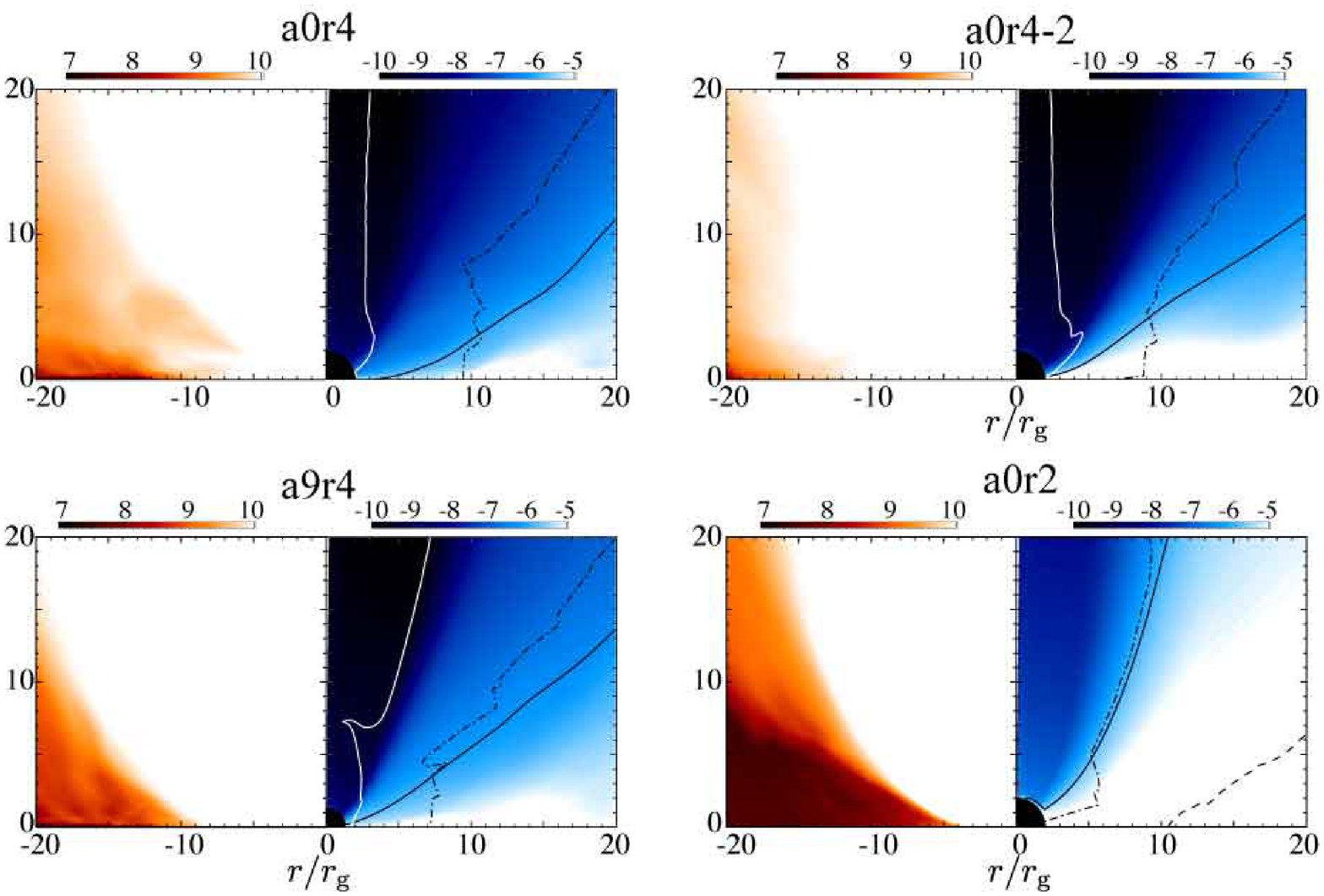}
  \caption{
The $\phi-$averaged density (right) and gas temperature (left) profiles
 for each models. Black solid lines indicate the photosphere,
 where $\tau_\mathrm{tot}=1$, and the effective optical depth
 is unity on the dashed lines.
 Dot-dashed lines show where the dynamical time is comparable to the
 Compton cooling time, so that
 the Compton cooling might effectively work below the lines. 
 White lines denote expected surface where the gas temperature
 decreases down to $10^{10}\ \mathrm{K}$ due to the Compton cooling.
}
\label{fig3}
\end{figure}

In Figure \ref{fig3} the $\phi-$averaged density 
and gas temperature profiles are plotted on $r-\theta$ plane. 
We find that 
the geometrical thickness of the high-density region 
(white region on the right)
at $r\lesssim 10r_{\rm g}$ is very large for model a0r2
and is relatively small for the other models.
This is caused by the difference of the mass accretion rate,
$\dot{M} \gg L_{\rm Edd}$ for model a0r2
and $\dot{M} \lesssim L_{\rm Edd}$ for the other models.
In this figure, we find 
the gas temperature tends to be high (low) 
in the low (high) density regions.
The temperature in the high-density regions is around $10^7$K,
and, on the other hand, is larger than $10^9$K in the less dense regions.

In this figure, the black solid lines show 
the photosphere ($\tau_\mathrm{tot}=1$), 
and the dashed lines mean the surface 
where the effective optical depth, $\tau_\mathrm{eff}$, becomes unity. 
Here these optical depths are calculated from the polar axis,
\begin{equation}
 \tau_\mathrm{tot}=
  \int_0^\theta \gamma \rho (\kappa_\mathrm{abs}+\kappa_\mathrm{sca})
  \sqrt{g_{\theta\theta}}d\theta',
\end{equation}
and 
\begin{equation}
 \tau_\mathrm{eff}=
  \int_0^\theta \gamma \rho
  \sqrt{(\kappa_\mathrm{abs}+\kappa_\mathrm{sca})\kappa_\mathrm{sca}} \sqrt{g_{\theta\theta}}d\theta',
\end{equation}
\citep{2014MNRAS.441.3177M}. 
In all models, 
the opening angle of the photosphere is not large,
$\sim 25^\circ$ for model a0r2 and $\sim 55^\circ$ for the other models.
Since the absorption opacity is 
much smaller than the scattering opacity,
the effective optical depth is less than the total optical depth.
Thus, the surface of $\tau_\mathrm{eff}=1$ appears deep 
inside the photosphere.
This figure also shows 
that the disk of $\tau_\mathrm{eff}>1$ is truncated 
around the black hole.
The truncation radius $r_\mathrm{tr}$, 
at which the surface of $\tau_\mathrm{eff}=1$ 
reaches to the equatorial plane, is 
$r_\mathrm{tr}\sim 30 r_{\rm g}$ for the low-$\rho_0$ models (a0r4, a0r4-2, a9r4),
and $r_\mathrm{tr}\sim 10 r_{\rm g}$ for the high-$\rho_0$ models (a0r2).
It implies that the inner part of the disk does not emit 
the blackbody radiation.

The seed photons emitted at the surface of 
$\tau_\mathrm{eff}=1$ suffer from the numerous scattering
before escaping from the photosphere.
The Compton upscattering would play an important role for
producing the hard X-ray photons,
since the gas is very hot ($\gtrsim 10^9$K) in the regions 
between the photosphere and the surface of $\tau_\mathrm{eff}=1$.
Hereafter we call this hot regions ``overheated regions''.
As we will show below, the gas in not in LTE and 
we find $T_{\rm gas} \gtrsim 10T_{\rm rad}$ in the overheated regions.

\begin{figure}[t]
 \includegraphics[width=8.5cm]{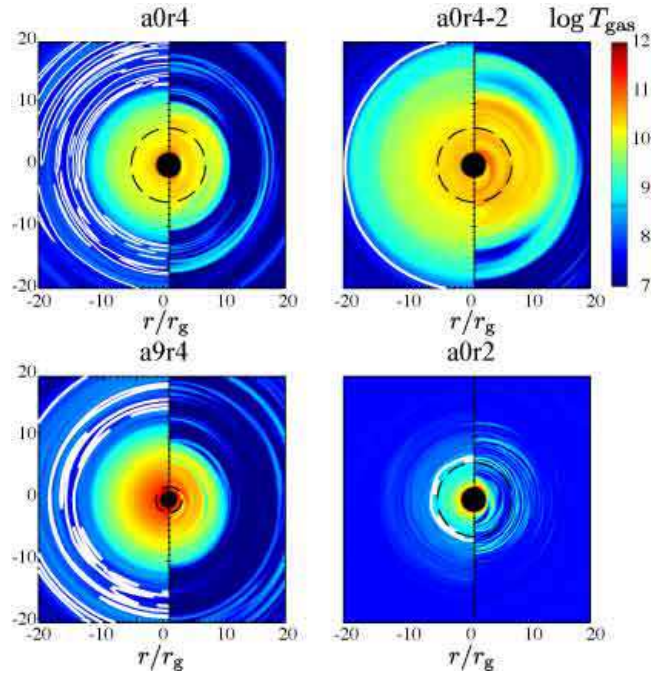}
  \caption{
 Gas temperature profile on the equatorial plane.
 In each panels, the time averaged $T_\mathrm{gas}$ 
 and $T_\mathrm{gas}$ at $t=t_\mathrm{end}$
 are plotted on the right and left, respectively.
 White lines represent where $T_\mathrm{gas}/T_\mathrm{rad} = 10$. 
 Black dashed lines indicate the ISCO.
}
\label{fig4}
\end{figure}

Figure \ref{fig4} shows the gas temperature distribution on the
equatorial plane. 
In each panels,
the time averaged $T_\mathrm{gas}$ is shown in the left half
and the right half of the panels indicates 
the snapshot at the end of the range of the time,
$t=0.15$ (a0r4-2 and a0r2)
and $t=0.3$ (a0r4 and a9r4).
Time averaged contours where $T_\mathrm{gas}/T_\mathrm{rad}=10$
are plotted by white solid lines
in the right half in each panels.
Black circles indicate an innermost stable circular orbit
(ISCO).

This figure shows that
the gas temperature increases 
as approaching to the black hole. 
Although the gas temperature is $\sim 10^7 \mathrm{K}$ 
for a larger radius (blue),
it exceeds $10^9 \mathrm{K}$ around the black hole
(green, yellow, and red).
This hot region corresponds to the overheated region
on the equatorial plane.
Although the line are complicated in the models of a0r4 and a9r4,
$T_{\rm gas}$ is comparable to or slightly larger than $T_{\rm rad}$
in the blue regions,
and is at least ten times larger than $T_{\rm rad}$ 
in the overheated regions.

In comparison with the model a0r4 (upper-left),
the gas temperature for the model a9r4 (lower-left)
is very high in the vicinity of the black hole.
Although the red region appears within $r \sim 7r_{\rm g}$
in the case of a9r4,
such a very hot region is not produced for the model a0r4.
In contrast, 
the gas temperatures in two models are approximately equal
in the regions of $r\gtrsim 10r_{\rm g}$.
These results imply the rotation of the black hole 
contributes to the heating of the gas around the black hole
(we will discuss later).
Also, the right panels indicate 
the overheated region shrinks as 
the accretion rate increases.
The diameter of the overheated region is about $14r_{\rm g}$
for the model a0r2 and $40r_{\rm g}$
for the model a0r4-2.
The overheated region extends to the outside of the ISCO
for low-$\rho_0$ models (a0r4, a0r4-2, a9r4).

\begin{figure}[t]
 \includegraphics[width=8.5cm]{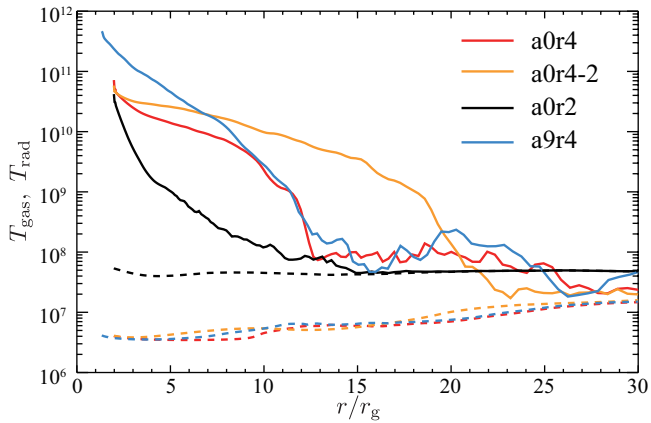}
  \caption{Radial profiles of $<T_\mathrm{gas}>_{\rho,\theta\phi}$
 (solid) and $<T_\mathrm{rad}>_{\rho,\theta \phi}$ (dashed).
 Red, orange, black and blue lines show
 results of models a0r4, a0r4-2, a0r2, and a9r4, respectively.
}
\label{fig5}
\end{figure}

The profiles of the gas and radiation temperatures are 
more clearly understood in Figure \ref{fig5},
where we plot $<T_\mathrm{gas}>_{\rho,\theta\phi}$ (solid) 
and $<T_\mathrm{rad}>_{\rho, \theta\phi}$ (dashed)
as a function of the radius.
Red, orange, black and blue lines show results of models a0r4, a0r4-2,
a0r2, and a9r4, respectively. 
It is found that 
the radiation temperature is insensitive to the radius.
For the model a0r2, 
we find $<T_\mathrm{rad}>_{\rho,\theta\phi} \sim 4\times 10^7 \rm K$.
The radiation temperature for the other models 
is slightly lower than that of a0r2.
The gas temperature is comparable to the radiation temperature
at $r \gtrsim 25 r_\mathrm{g}$ for low-$\rho_0$ models 
(a0r4, a9r4, a0r4-2)
and at $r\gtrsim 10 r_\mathrm{g}$ for a high-$\rho_0$ model (a0r2).
As approaching the black hole, 
the gas temperature steeply increases
and deviates from the radiation temperature.
The radius of the overheated region ($r_{\rm over}$), 
at which $T_{\rm gas}=10T_{\rm rad}$,
is about $15r_{\rm g}$ for models a0r4 and a9r4,
$20r_{\rm g}$ for model a0r4-2,
and $7r_{\rm g}$ for model a0r2.

As we have already mentioned above,
the radius of the overheated region depends on the mass accretion rate.
Comparing models a0r2 ($\dot{M}\sim 430L_{\rm Edd}$) 
and a0r4-2 ($\dot{M}\sim 4.3L_{\rm Edd}$), 
we find $r_{\rm over}$ for a0r2 is less than that for a0r4-2.
This figure also shows that the gas temperatures 
is insensitive to the rotation of the black hole
at $r\gtrsim 10r_{\rm g}$ (see a0r4 and a9r4).
However, the gas is more effectively heated up in the case of a9r4.
In this model, the maximum temperature exceeds $10^{11}\rm K$
near the black hole,
and is 10 times larger than that of a0r4.
Although the gas temperature might be overestimated
since the Compton cooling is not taken into consideration in 
the present simulations,
our results indicate that the rotation of the black hole
contributes to the heating of the gas 
in the vicinity of the black hole.

So far, we have discussed 
with using the time- and/or angle-averaged structures.
However, our three-dimensional simulations 
show a non-axisymmetric structure, which are 
smeared out by taking the average.
The right half of each panels in Figure \ref{fig4}
is the snapshot of the gas temperature.
We can see patchy structure and/or spiral hot regions in all models.
Such a non-uniform structure would impact on the variability 
of observed spectra \citep{2015ApJ...799....1C}. 

\subsection{formation mechanisms of overheated region}
\label{macha}
\begin{figure}[t]
 \includegraphics[width=8.5cm]{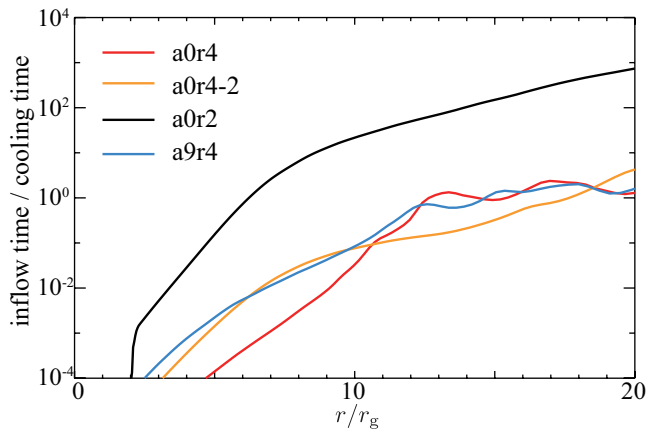}
  \caption{
 Radial profiles of the ratio of inflow time and cooling time.
 The inflow time is calculated by the equation (\ref{eq:dyn}) and 
 the cooling time is estimated based on the free-free emission.
 Red, orange, black and blue lines show
 results of models a0r4, a0r4-2, a0r2, and a9r4, respectively.}
\label{fig6}
\end{figure}

Figure \ref{fig6} shows the ratio of the inflow time, $t_\mathrm{in}$, 
and the cooling time, $t_\mathrm{cool}$, 
for models a0r4 (red solid), 
a0r4-2 (orange solid), 
a0r2 (black solid), 
and a9r4 (blue solid). 
These time scales are evaluated as
\begin{equation}
 t_\mathrm{in}(r) =-\int_{r_{\mathrm{H}}}^r
  \frac{dr'}{<v_r>_\mathrm{\rho,\theta\phi}},
\end{equation}
\citep{2014MNRAS.441.3177M}, and 
\begin{equation}
 t_\mathrm{cool} = 
\left< \frac{e}{4\pi \kappa_\mathrm{abs}\mathrm{B}}
\right>_\mathrm{\rho,\theta\phi}.\label{eq:bremss}
\end{equation}
We can see that the $t_\mathrm{in}/t_\mathrm{cool}$ is larger for a
larger radius. 
Close to the black hole, the inflow speed increases with decreasing
radius, so that $t_\mathrm{in}$ is reduced. In addition, the
cooling time becomes longer for the inner region since the mass density
decreases. 
As a consequence, the gas accretes onto the black hole without cooling,
producing the overheated region \citep{1998MNRAS.297..739B}.  

Why does the truncation radius shift inward
as the mass accretion rate increases?
This is simply understood by the density dependence of
the cooling time.
Since we employ the free-free absorption opacity,
$\rho \kappa_{\rm abs} \propto \rho^2$, in the present work,
the cooling time is in inverse proportion to
the density, 
$t_\mathrm{cool} [\propto e/(\rho \kappa_\mathrm{abs})] \propto \rho^{-1}$.
The larger $\rho_0$ is,
the larger the density of the disk and the mass accretion rate becomes.
Thus, the overheated region shrinks as the mass accretion rate increases.

Next, we discuss about the reason why 
the maximum gas temperature in the overheated region 
is higher for the case of the rotating black hole (a9r4) 
than for the case of the non-rotating black hole (a0r4).
The one of the most plausible mechanism is that 
the rotational energy of the rotating black hole 
is transmitted to the matter around the black hole.
When the black hole threaded by the magnetic field rotates, 
its rotational energy is extracted through the magnetic field
\citep{1977MNRAS.179..433B, 2004ApJ...611..977M, 2008PhRvD..78b4004T}.
This process enhances the energy of the black hole magnetosphere 
and induces the launching of the jets
\citep{2006MNRAS.368.1561M,2010MNRAS.408..752P,2012JPhCS.372a2040T,2013Sci...339...49M}.

\begin{figure}[t]
 \includegraphics[width=8.5cm]{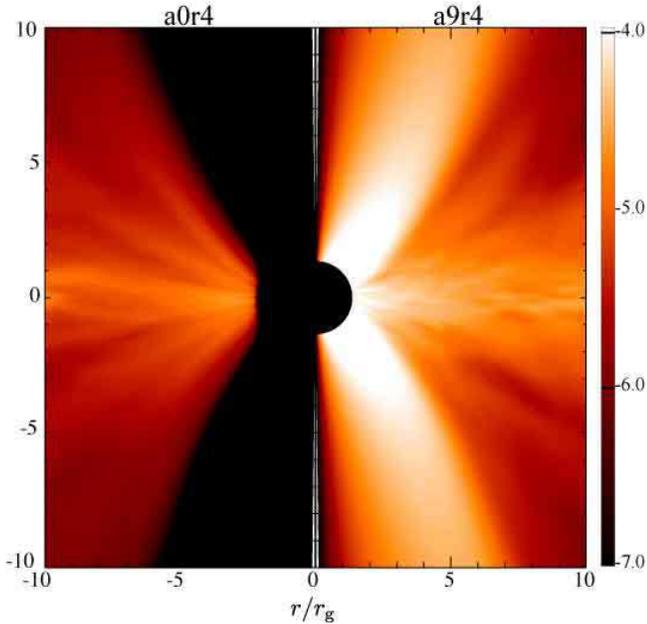}
  \caption{
 Profiles of the outward Poynting flux $<-M_t^r>_{1,\phi}$ 
 around the non-rotating black hole (a0r4) and the rotating black hole
 (a9r4).}
\label{fig7}
\end{figure}

Figure \ref{fig7} shows the radial component of the Poynting flux $<-M_t^r>_{1,\phi}$
normalized by $\rho_0$ for model a0r4 (left) and a9r4 (right). 
It is found in this figure that 
the Poynting flux is outward in most of the region.
Also, we find that the electromagnetic energy
is more effectively transported outward 
for the rotating black hole than
for the non-rotating black hole.
This is conspicuous near the black hole.
Indeed, white regions appear only around the rotating black hole 
(see right panel).
For the rotating black hole case, 
the strong Poynting flux is mainly emitted to the direction of 
$\theta \sim \pi/4$ and $3\pi/4$.
In addition, we find the substantial Poynting flux 
in the direction along the equatorial plane, $\theta \sim \pi/2$.
The electromagnetic energy transported via the Poynting flux 
would work to heat up the matter in the overheated regions.

\begin{figure}[t]
 \includegraphics[width=8.5cm]{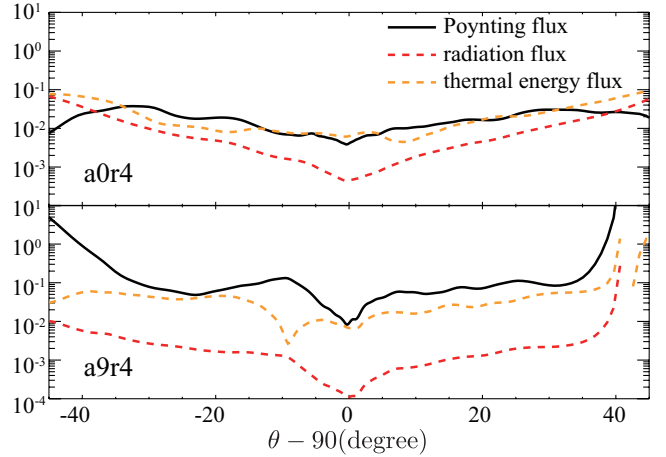}
  \caption{Radial component of the energy fluxes 
 at $r=5r_{\rm g}$ for a0r4 (top) and a9r4 (bottom). 
 Black solid lines show
 the outward Poynting flux, while orange and red dashed lines show
 the inward thermal energy flux and the radiation flux, respectively.}
\label{fig8}
\end{figure}

In Figure \ref{fig8},
we plot 
the outward Poynting flux, $<-M_t^r>_{1,\phi}$ (black solid),
inward radiation flux, $<R_t^r>_{1,\phi}$ (red dashed),
and inward thermal energy flux,
$<\Gamma p_\mathrm{gas} u_t u^r/(\Gamma-1)>_{1,\phi}$ 
(orange dashed) at $r=5r_\mathrm{g}$
as a function of $\theta$.
Top and bottom panels show results for a0r4 and a9r4. 
This figure shows 
that the gas energy flux dominates the radiation energy flux
in both models, since the gas is overheated and its
temperature much exceeds the radiation temperature
at around $r=5r_\mathrm{g}$ (see Figures \ref{fig4} and \ref{fig5}).

As shown in this figure,
the outward Poynting flux becomes larger for the rotating black hole
case than that for the non-rotating black hole case.
This is because that the rotational energy is extracted 
from the rotating black hole through the magnetic field.
The outward Poynting energy flux should increase
as increasing the black hole spin $a^*$.
Since the electromagnetic energy 
is dissipated around the black hole,
the gas temperature in the overheated region
becomes higher in the case of the rotating black hole.
We note that the ideal magnetohydrodynamics is assumed in the 
present simulations, so that 
the dissipation of the magnetic energy 
might be originated from the numerical resistivity.

Another possible mechanism of the energy dissipation is the mode conversion of the MHD waves. 
The Alfv\'en waves excited around the black hole propagates inside the
accretion disks. Since the density contrast exists inside the disks, 
Alfv\'en waves would suffer from the mode conversion to the fast and
slow modes due to the non-linear effects
\citep{1978ApJ...219..700G,1986JGR....91.4171T}. The generated
compressional waves would heat up the gas \citep{2012ApJ...755...76T}. 


\subsection{compton cooling}
\label{comp}
In the present simulations, the Compton cooling is not included.
The Compton upscattering works to produce the high-energy photons,
and at the same time induces the cooling of the gas.
On the dot-dashed lines in Figure \ref{fig3},
the dynamical timescale,
\begin{equation}
 t_\mathrm{dyn}=\left<\frac{r}{v_r}\right>_{1,\phi},
  \label{eq:dyn}.
\end{equation}
equals to the Compton cooling timescale,
\begin{equation}
 t_\mathrm{comp}=\left<-\frac{e}
  {\rho \kappa_\mathrm{sca} \hat E_\mathrm{rad}
  \frac{4 k_\mathrm{B}
  (T_\mathrm{exp} - T_\mathrm{rad})
  }{m_\mathrm{e}}
  }\right>_{1,\phi},\label{eq:tcompton}
\end{equation}
where $T_\mathrm{exp} =10 T_{\rm rad}$.
Thus the $t_\mathrm{comp}$ indicates the time scale that 
the gas temperature decreases to $T_\mathrm{gas}=T_\mathrm{exp}$.
Here note that the result does not change
even if we set $T_\mathrm{gas}$ to be $5T_{\rm rad}$.
%

Below the dot-dashed lines,
the gas is cooled by the Compton cooling.
In contrast, the gas is overheated above the lines,
$T_{\rm gas} > 10 T_{\rm rad}$.
That is, even if the Compton cooling effectively works, 
the hard X-ray photons are produced by 
the Compton upscattering in the regions 
between the solid line and the dot-dashed line
(narrowed overheated regions).
The radius of the narrowed overheated region 
near the equatorial plane is $\sim 10r_{\rm g}$ for models a0r4, a9r4, 
and a0r4-2.
For the model of a0r2, the narrowed overheated regions does not appear
on the equatorial plane.
Although the Compton cooling also reduces the truncation radius,
the disk truncation does not disappear.
We find $r_{\rm tr} \sim 10r_{\rm g}$ (a0r4, a9r4, and a0r4-2)
and $r_{\rm tr} \sim 4r_{\rm g}$ (a0r2).

In Figure \ref{fig3}, the gas temperature is much higher than
$10^{10} \mathrm{K}$ in wide regions (white in the left panels). 
However, the gas temperature in this region would be less than $10^{10}
\mathrm{K}$ in reality, except for near the polar axis for models a0r4
a0r4-2, and a9r2. This is due to the Compton cooling. 
The white lines in Figure \ref{fig3} show where the dynamical
time (equation 24) becomes comparable to the cooling time, which is
estimated by setting $T_\mathrm{exp}=10^{10} \mathrm{K}$ in equation
(25). 
For models a0r4, a0r4-2 and a9r4, we find that the gas temperature is 
thought to be $\gtrsim 10^{10} \mathrm{K}$ at the vicinity of the polar
axis (between the rotation axis and the white line), while it is
expected 
to be $\lesssim 10^{10} \mathrm{K}$ in the outer region. 
We also note that the region where $T_\mathrm{gas}>10^{10} \mathrm{K}$
disappears for the high $\dot M$ case (a0r2). Our results are consistent
with \cite{2015MNRAS.454.2372S}, in which the very hot regions do not form
via the Compton cooling for the case of $\dot M \gtrsim 100 L_\mathrm{Edd}$.

We can conclude that,
even if the Compton cooling effectively decreases the gas temperature,
the disk of $\tau_{\rm eff}>1$ is truncated around the black hole 
and is sandwiched by the overheated regions 
(between the solid line and the dot-dashed line in Figure \ref{fig3}),
enhancing the hard X-ray spectra 
via the Compton upscattering in the overheated regions.

\section{Summary \& Discussion}
By performing three-dimensional GRRMHD simulations
of accretion flows around the black holes,
we revealed that 
the relatively cold disk, of which 
the gas and the radiation temperatures are 
$\sim {\rm several} \times 10^{7}$K
and the effective optical depth is $>1$,
is truncated at around the black hole.
The hot and rarefied regions (overheated regions),
in which the gas temperature ($T_{\rm gas} >10^9 {\rm K}$)
is more than ten times higher than
the radiation temperature, 
appear within the truncation radius and above the cold disk.
The cold disk is sandwiched by the overheated regions,
so that the hard X-ray photons would be produced by the 
Compton upscattering in the overheated regions.
%
The truncation radius decreases with an
increase of the mass accretion rate, 
since the cooling is effective in the dense disk.
In the present simulations, it is 
$\sim 30 r_{\rm g}$ for the case of $\dot{M} \sim L_{\rm Edd}$ 
and $\sim 10 r_{\rm g}$ for the case of
$\dot{M} \sim 100 L_{\rm Edd}$.
Although the overheated regions shrink 
if the Compton cooling effectively reduces the gas temperature,
the sandwich-structure does not disappear
at the range of $\dot{M}\sim 1-100 L_{\rm Edd}$.
The maximum gas temperature in the overheated region
is about ten times higher for the rotating 
black hole than for the non-rotating black hole.
In the case of the rotating black hole,
since the electromagnetic energy is enhanced 
at the very vicinity of the black hole and
transported outward by the Poynting flux, 
the matter would be effectively heated up.

Our low-$\rho_0$ models would explain the very high state of the BHBs.
The very high state is thought to appear 
for the near- or sub-Eddington case, 
and the power-low component is dominant over the soft component.
\citet{2004MNRAS.353..980K} reported that 
the inner part of the optically-thick disk does not reach 
to the ISCO radius in this state,
and the effective Compton upscattering produces the strong power-low spectra.
Such features nicely fit our results of models a0r4, a0r4-2, and a9r4,
whereby the truncated disk is sandwiched by the overheated regions.
On the other hand, 
our model a0r2 would correspond to the slim disk state.
\cite{2008PASJ...60..653V} succeeded in reproducing the observed X-ray
spectra of the ultra luminous X-ray source, 
using the slim disk model which is not truncated.
In our super-Eddington model (a0r2), the truncation radius is very 
close to the ISCO radius, so that our result is not inconsistent with 
the observations.
%
\begin{deluxetable}{lccc}[t]
\tabletypesize{\scriptsize}
\tablecaption{typical radius obtained by simulations}
\tablewidth{0pt}
\tablehead{
\colhead{model} 
 & \colhead{inflow-outflow equilibrium} 
 &\colhead{$r_\mathrm{over}$}
 &\colhead{$r_\mathrm{tr}$} 
}
\startdata
a0r4   &  $\sim 15 $ & $\sim 15(10)$ & $\sim 30(9)$ \\
a9r4   &  $\sim 15 $ & $\sim 15(10)$ & $\sim 30(7)$\\
a0r4-2 &  $\sim 20 $ & $\sim 20(10)$ & $\sim 30(9)$\\
a0r2   &  $\sim 20 $ & $\sim 7(2)  $ & $\sim 10(4) $ 
\enddata
\tablecomments{From left to right, model, the size of the inflow-outflow
 equilibrium, the equatorial radius of the overheated region, and the
 equatorial radius where $\tau_\mathrm{eff}=1$. The value denoted in the
 bracket is obtained by taking into account the Compton cooling. These
 radii are denoted in the unit of $r_\mathrm{g}$ }
\label{tab1}
\end{deluxetable}

Here we note that,
the multi-frequency radiation transfer calculations are needed 
to accurately investigate the emergent spectra.
Such calculations are attempted by \cite{2012ApJ...752...18K}
in which the hard X-ray spectra is obtained by performing 
the post-processing radiation transfer calculations
including the Compton upscattering and down-scattering
\citep[see][]{2015MNRAS.451.1661Z, 2015arXiv151004208N}.
However, it is difficult to accurately treat
the gas-radiation interaction via the Compton process
in the hydrodynamics simulations.
In \cite{2009PASJ...61..769K}, the Compton heating/cooling is 
calculated by solving the Kompaneets equation 
based on the assumption that 
the radiation has a blackbody spectrum.
Recently, \cite{2015MNRAS.454.2372S} also performed hydrodynamics
simulations taking into account the Comptonization.
In their method,
the assumption of the blackbody spectrum is relaxed 
and the conservation of number of photons is carefully treated.
However, as long as the Kompaneets equation is employed,
the simulations would induce some inaccuracies 
in the regions of anisotropic radiation fields 
like around the photosphere.
Multi-frequency RMHD simulations are 
necessary to resolve this problem,
but such simulations are too time consuming to perform.

The time for simulations is limited in the present work
as shown in Figure \ref{fig1}, especially for run C.
The structure of the magnetic fields might change by graduation
due to the magnetic dynamo.
If it does, the structure of the overheated regions is influenced. 
Thus, the long-term three-dimensional simulations 
should be explored in future work,
since it is well known that no magnetic dynamo works
in the axisymmetric calculations.
Two-dimensional simulations including the sub-grid dynamo model 
is another way \cite{2015MNRAS.447...49S}.

Simulations of higher resolution are left as important future work. In
order to revolve the magnetorotational instability, it has been
reported that $Q_\theta \gtrsim 6$ is required \citep{2014MNRAS.441.3177M}. 
Here, $Q_\theta$ is evaluated as $Q_\theta=\lambda_\mathrm{MRI} / r
d\theta$ with $d\theta$ being the grid size in theta direction
\citep{2013ApJ...772..102H,2014MNRAS.441.3177M}.
In model a0r4, in which the disk is the thinnest in the present work, 
$<Q_\theta>_{\rho,\theta \phi}$ is $\simeq 7$ around
$r=10r_\mathrm{g}$. 
This value is comparable to the required value of $6$, so that it
is better to perform simulations with small $d\theta$.


Finally, M-1 closure method employed in the present work
is known to be somewhat problematic 
in the optically very thin or moderately thin regions.
The accurate radiation fields can be obtained 
by solving radiation transfer equations.
Such a method has been proposed by 
\cite{2014ApJS..213....7J, 2015ApJ...807...31R,2016ApJ...818..162O}

\acknowledgments
Numerical computations were carried out on Cray XC30 at the Center for
Computational Astrophysics of National Astronomical Observatory of
Japan, on FX10 at Information Technology
Center of the University of Tokyo, and on K computer at AICS.
This work is supported in part by JSPS Grant-in-Aid for 
Scientific Research (C) (15K05036 K.O., 15H00782 Y.S.). 
A part of this research has been funded by MEXT HPCI STRATEGIC PROGRAM 
and the Center for the Promotion of Integrated Sciences (CPIS) of Sokendai.


\end{document}